\documentclass[aps,prb,twocolumn,preprintnumbers,amsmath,amssymb,superscriptaddress]{revtex4}%
\usepackage{graphicx}
\usepackage{dcolumn}
\usepackage{amsmath}
\usepackage{color}
\usepackage{amsfonts}
\usepackage{amssymb}%
\providecommand{\U}[1]{\protect\rule{.1in}{.1in}}
\begin{document}
\title{Anomalous gap ratio in anisotropic superconductors: aluminum under pressure}
\author{Rustem~Khasanov}
\email{rustem.khasanov@psi.ch}
\affiliation{Laboratory for Muon Spin Spectroscopy, Paul Scherrer Institut, CH-5232
Villigen PSI, Switzerland}
\author{Igor I. Mazin}
\affiliation{Department of Physics and Astronomy, George Mason University, Fairfax,
Virginia 22030, USA}
\affiliation{Quantum Science and Engineering Center, George Mason University, Fairfax,
Virginia 22030, USA}

\begin{abstract}
Pressure dependence of the thermodynamic critical field $B_{\mathrm{c}}$ in
elemental aluminum was studied by means of the muon-spin rotation/relaxation
technique. Pressure enhances the deviation of $B_{\mathrm{c}}(T)$ from the
parabolic behavior, expected for a typical type-I superconductor, thus
suggesting the weakening of the gap ratio $\langle\alpha\rangle=\langle
\Delta\rangle/k_{\mathrm{B} }T_{\mathrm{c}}$ ($\langle\Delta\rangle$ is the
average value of the superconducting energy gap, $T_{\mathrm{c}}$ is the
transition temperature and $k_{\mathrm{B}}$ is the Boltzmann constant). With
the pressure increase from 0.0 to $\simeq1.6$~GPa, $\langle\alpha\rangle$
decreases almost linearly from 1.73 to 1.67. Our results imply, therefore,
that in elemental aluminum the gap ratio $\langle\alpha\rangle$ is smaller
than the weak-coupled BCS prediction $\alpha_{\mathrm{BCS}}\simeq1.764$ and it
is even further reduced under pressure.

\end{abstract}
\maketitle

%\pacs{74.70.Xa, 74.25.Bt, 74.45.+c, 76.75.+i}

In superconducting materials the coupling strength is generally established by
comparing the reduced gap (the ratio of the energy gap $\Delta$ to the
transition temperature $T_{\mathrm{c}}$) with the universal weak-coupling BCS
number $\alpha_{\mathrm{BCS}}=\Delta/k_{\mathrm{B} }T_{\mathrm{c}}%
=e^{\gamma_{\mathrm{E}}}/\pi\simeq1.764$ ($\gamma_{\mathrm{E}}$ is the Euler
constant, and $k_{\mathrm{B}}$ is the Boltzmann constant).\cite{BCS_PR_1957,
Tinkham_book_1975, Poole_Book_2014} Based on such comparison, the
superconductors are divided into the strong-coupled ($\alpha\gg1.764$),
intermediate-coupled ($\alpha\gtrsim1.764$) and  weak-coupled
($\alpha\simeq1.764$) classes. This division has a profound physical meaning:
the universal value $\alpha_{\mathrm{BCS}}$ corresponds to the temperature at
which the order parameter is destroyed by its own thermal fluctuations. Any
additional pair-breaking agent can only suppress $T_{\mathrm{c}}$ and, thus,
increase $\alpha$ above the $\alpha_{\mathrm{BCS}}$ value.

More accurate than BCS, the Eliashberg theory, accounts for the fact that
while virtual phonons bind electrons into Cooper pairs, the real (thermal)
phonons break them. In the weak-coupling limit, by definition, $T_{\mathrm{c}%
}$ is exponentially smaller than the energy of the pairing bosons, so the
concentration of thermal excited bosons, interacting with superconducting
electrons, is exponentially low. Consequently, the reduced gap stays
exponentially close to $\alpha_{\mathrm{BCS}}$. As coupling becomes stronger,
$T_{\mathrm{c}}$ becomes larger, compared to the boson frequencies, and the pair-breaking effect of real phonons is not negligible anymore. With
temperature, their number, and, correspondingly, their pair breaking effect
grows, and superconductivity becomes destroyed at $T_{\mathrm{c}%
}<T_{\mathrm{c,BCS}}$. Thus, the reduced gap $\alpha=\Delta/k_{\mathrm{B}%
}T_{\mathrm{c}}>\Delta/k_{\mathrm{B}}T_{\mathrm{c,BCS}}=\alpha_{\mathrm{BCS}}%
$. By modifying the phonon spectrum, \textit{i.e.} by adding very soft
phonons, one can generate a paradoxical situation, where superconductivity is
physically from weak-coupling (the phonons generating $T_{\mathrm{c}}$ are
exponentially harder than the gap), and yet have a huge $\alpha$
ratio.\cite{rot} One may also consider two different type of bosons, one
pairing and the other pair breaking even in virtual diagrams (for instance,
the former may be phonons and the latter spin fluctuations),\cite{magnons}
albeit, generally speaking, such a situation should also lead to an
enhancement of $\alpha.$

An opposite effect, a phonon spectrum that would render $\alpha<$
$\alpha_{\mathrm{BCS}},$ appears, on the first glance, impossible (barring a
dramatic change of the actual phonon spectrum between $T=0$ and $T=T_{c})$.
Note however, that
%one should keep in mind that
the above consideration
implicitly assumes the superconducting state characterized by one
single order parameter,  single in the sense that there is one number for all
Cooper pairs in the system. This assumption is, of course, violated in such
cases as multiband and(or) anisotropic superconductivity,
where the order parameters vary over the Fermi surface. In that case,
$T_{\mathrm{c}}$ is uniquely defined, but the way how one collapses the
function $\Delta({\mathbf{k}})$ to a single average number $\langle
\Delta\rangle$ is ambiguous. Interestingly, some most natural definitions of
$\langle\Delta\rangle$, resulting from experimental methods of measuring the
order parameter, may actually render the apparent $\langle\alpha
\rangle=\langle\Delta\rangle/k_{\mathrm{B}}T_{\mathrm{c}}$ to be smaller than
$\alpha_{\mathrm{BCS}}$.

The elemental aluminum is often cited as a canonical weak-coupling
superconductor. Indeed, the tunneling experiments of Blackford and
March,\cite{Blackford_CJP_1968} performed on Al polycrystalline films resulted
in $\alpha=1.765(5)$, in excellent agreement with the BCS prediction. At the
same time, experiments on high-quality aluminum single crystals revealed that
the superconducting gap is anisotropic, \textit{i.e.} it has different value
along different crystallographic directions. The $\alpha$ value was found to
vary from $\simeq1.5$ to $\simeq1.8$,\cite{Wells_PR_1970, Blackford_JLTP_1976,
Kogure_JPSJ_1985} around the Fermi surface. This suggests, after a closer
look, that the average reduced gap $\langle\Delta\rangle$ (of course, depending on the averaging
protocol) may become smaller than the universal BCS
limit.\cite{Clem_AnnPhys_1966, Clem_PR_1967}

The first attempts to evaluate gap anisotropy of superconducting aluminum
theoretically go back to 1971.\cite{Dynes_Physica_1971,Leavens_AnnPhys_1972}
These were using empirical model pseudopotentials and a spherical Fermi
surface, and were later somewhat improved to include a realistic Fermi
surface.\cite{Leug_JLTP_1976} Meza-Montes \textit{et al.}%
\cite{Meza-Montes_JLTP_1988} improved that further by taking into account the
anisotropy of the Coulomb pseudopotential $\mu^{\ast},$ albeit they never took
into account the anisotropy of the logarithmic renormalization of $\mu^{\ast}%
$, even though it can be argued to be more important.\cite{us} While these
calculations may be considered subpar by modern standards, they agree
qualitatively with the experimental observations of
Refs.~\onlinecite{Wells_PR_1970, Blackford_JLTP_1976, Kogure_JPSJ_1985} (see
also a cummulative table in Ref.~\onlinecite{rev}). It is interesting to note,
that one of the early theoretical predictions, which is not
verified experimentally so far, suggest that the gap anisotropy in aluminum should
greatly increase with the applied pressure $p$.\cite{Leavens_CanJPhys_1972}

In this paper we present experimental evidence that the reduced thermodynamic
gap in elemental aluminum, defined as the gap corresponding to the
thermodynamic critical field, $\left\langle \Delta\right\rangle =B_{\mathrm{c}%
}(0)/\sqrt{4\pi N(0)}$ [$N(0)$ is the density of states at the Fermi level and
$B_{\mathrm{c}}(0)=B_{\mathrm{c}}(T=0)$ is the zero-temperature value of the
thermodynamic critical field], is \textit{smaller} than $\alpha_{\mathrm{BCS}%
}$ and it \textit{decreases} under pressure. The analysis of the experimental
$B_{\mathrm{c}}(T,p)$ dependencies within the anisotropic gap model of
Clem,\cite{Clem_AnnPhys_1966, Clem_PR_1967} suggest that the mean-squared
anisotropy value
\begin{equation}
\langle a^{2}\rangle\equiv\frac{\lbrack\Delta({\mathbf{k}})-\langle
\Delta\rangle]^{2}}{\langle\Delta\rangle^{2}} \label{eq:anisotropy-squared}%
\end{equation}
changes, as the pressure increases from $p=0.0$ to $\simeq1.6$~GPa, from
$\langle a^{2}\rangle\simeq0.013$ to $\simeq0.035$, which, in turn, leads to a
decrease of $\langle\alpha\rangle$ from 1.73 to 1.67.

%%%%experimental

The description of cylindrically shaped Al samples, piston-cylinder type of pressure cells, and
transverse-field muon-spin rotation/relaxation (TF-$\mu$SR) under pressure
experiments are given in the Supplementary part of the
manuscript.\cite{Supplemental_part}

%The Al samples were prepared from commercially available 5~mm in diameter aluminium rod (99.999\% purity). Pieces of the 'soft' Al rod were pressed into  a cylindrically-shaped samples with the diameter and the height $d_{1}=7.9$, $h_{1}\simeq14$~mm and $d_{2}=6.9$, $h_{2}\simeq13.0$~mm for the sample \#1 and \#2, respectively. The samples were placed inside the single-wall (8~mm inner diameter, maximum pressure $p_{\mathrm{max}}\simeq1.4$~GPa) and the double-wall (7~mm inner diameter, $p_{\mathrm{max}}\simeq2.0$~GPa) pressure cells made of MP35N alloy. The construction of the pressure cells is similar to the one described in Refs.~\onlinecite{Khasanov_HPR_2016,Shermadini_HPR_2017}. The muon-spin rotation/relaxation ($\mu$SR) under pressure experiments were performed at the $\mu$E1 beamline by using GPD spectrometer (Paul Scherrer Institute, PSI Villigen,  Switzerland).\cite{Khasanov_HPR_2016} The $^{4}$He cryostat equipped with the $^{3}$He inset was used. The external magnetic field $B_{\mathrm{ex}}$ was applied perpendicular to the initial muon-spin direction, which corresponds to the transverse-field (TF-$\mu$SR) geometry. The experiments were performed in the temperature range of 0.25 to 1.5~K and in the field range of 0.5 to 15 mT.

\begin{figure}[tbh]
%\centering
\includegraphics[width=0.9\linewidth]{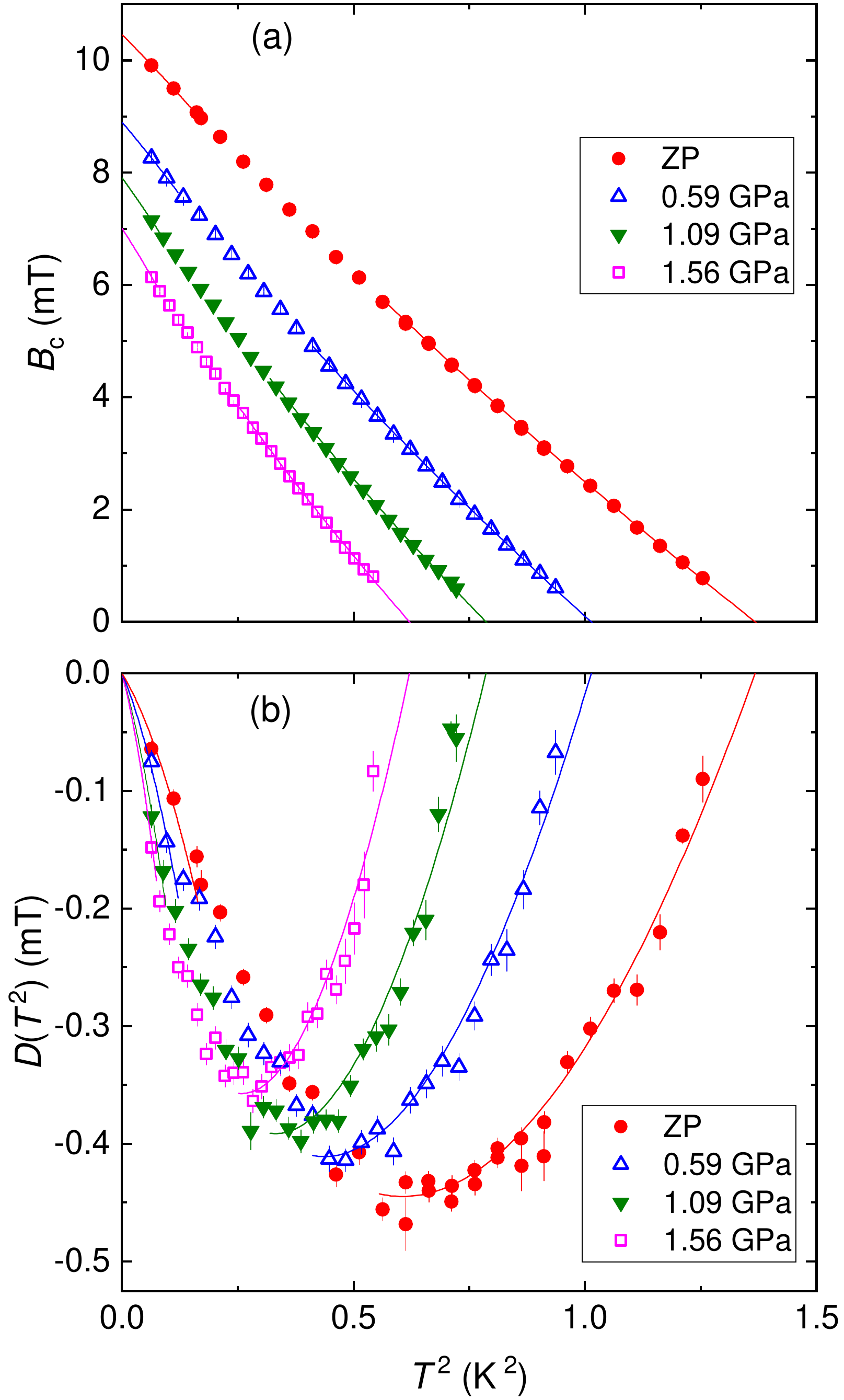}
%\vspace{-1.0cm}
\caption{(a) Temperature dependencies of the thermodynamic critical field
$B_{\mathrm{c}}$ of elemental aluminum measured at pressures $p=0.0$, 0.59,
1.09, and 1.56 GPa. The solid lines are fits of Eqs.~\ref{eq:Bc_low-T}, \ref{eq:Bc_high-T} to the low-temperature [$(T/T_{\mathrm{c}})^{2}<0.1$]
and the high-temperature [$0.5<(T/T_{\mathrm{c}})^{2}<1$] parts of $B_{\rm c}%
(T^{2},p)$ data, respectively. (b) Temperature dependencies of the deviation
function $D(T^{2})=B_{\mathrm{c}}(T^{2})-B_{\mathrm{c}}(0)\cdot
[1-(T/T_{\mathrm{c}})^{2}]$ at pressures $p=0.0$, 0.59, 1.09, and 1.56 GPa.
The solid lines are fits of Eqs.~\ref{eq:D_low-T}, \ref{eq:D_high-T} to
the low-temperature and the high-temperature parts of $D(T^{2},p)$ data, respectively.}%
\label{fig:Bc_Deviation}%
\end{figure}

The TF-$\mu$SR measurements were performed in the intermediate state of
superconducting aluminum, \textit{i.e.} when the sample volume is separated
into the normal state and the superconducting (Meissner)
domains.\cite{Tinkham_book_1975, Poole_Book_2014, deGennes_Book_1966,
Kittel_Book_1996, Prozorov_PRL_2007, Prozorov_NatPhys_2008,
Khasanov_Bi-II_PRB_2019, Karl_PRB_2019, Khasanov_Ga-II_PRB_2020,
Khasanov_AuBe_PRR_2020} The external field $B_{\mathrm{ex}}$ was applied
perpendicular to the sample's cylindrical axis. In this geometry the
demagnetization factor was estimated to be $n\simeq
0.42$,\cite{Supplemental_part, Prozorov_PRAppl_2018} so the intermediate state was
set in the region of $B_{\mathrm{c}}<B_{\mathrm{ex}}%
\lesssim0.58B_{\mathrm{c}}$. The modified $B-T$-scan measuring scheme, as
discussed in Refs.~\onlinecite{Karl_PRB_2019, Khasanov_AuBe_PRR_2020}, was
used. At each particular temperature the measured points were reached by first
decreasing $B_{\mathrm{ex}}$ to zero and then increasing it back to the
measuring ones. The $B-T$ points were taken along $\simeq0.9$, 0.825, 0.75,
and $0.675B_{\mathrm{c}}(T)$ lines by considering the $B_{\mathrm{c}}(p,T)$
curves as they were determined in
Refs.~\onlinecite{Boughton_book_1970, Palmy_Physica_1971}. Such a procedure
allows us to exclude effects of `supercooling',\cite{Cochran_PR_1956,
Faber_PRSL_1957, Saint-James_PhyLett_1963, McEvoy_SSC_1967, Park_SSC_1967,
Feder_SSC_1967, Smith_SSC_1968, Smith_PhCondMat_1970, Egorov_PRB_2001} which
were found to be particularly strong in the elemental aluminum
superconductor.\cite{Cochran_PR_1956, Faber_PRSL_1957, Smith_PhCondMat_1970}

The temperature dependencies of the thermodynamic critical field
$B_{\mathrm{c}}$ at pressures ranging from $p=0.0$ to $\simeq1.56$~GPa are
presented in Fig.~\ref{fig:Bc_Deviation}~(a). The magnetic field distribution
in a type-I superconductor in the intermediate state, which is probed directly
by means of TF-$\mu$SR, consists of two peaks corresponding to the response of
the domains remaining in the Meissner state ($B=0$) and in the intermediate
state ($B\equiv B_{\mathrm{c}}>B_{\mathrm{ex}}$). Consequently, in TF-$\mu$SR
experiments the value of $B_{\mathrm{c}}$ is directly and very precisely
determined by measuring the position of $B>B_{\mathrm{ex}}$
peak.\cite{Khasanov_Bi-II_PRB_2019, Karl_PRB_2019, Khasanov_Ga-II_PRB_2020,
Khasanov_AuBe_PRR_2020, Egorov_PRB_2001, Leng_PRB_2019,
Gladisch_HypInteract_1979, Grebinnik_JETP_1980, Beare_PRB_2019,
Kozhevnikov_JSNM_2020} The raw TF-$\mu$SR data and details of the data
analysis procedure, are presented in the Supplementary part of the
manuscript.\cite{Supplemental_part} Deviations of the $B_{\mathrm{c}}$
\textit{vs.} $T$ curves from the parabolic function: $D(T^{2})=B_{\mathrm{c}%
}(T^{2})-B_{\mathrm{c}}(0)[1-(T/T_{\mathrm{c}})^{2}]$ are shown in
Fig.~\ref{fig:Bc_Deviation}~(b). Following
Refs.~\onlinecite{Clem_PR_1967, Clem_AnnPhys_1966, Harris_PR_1968, Padamsee_JLTP_1973, Johnston_SST_2013, Khasanov_AuBe_PRB_2020}
the shape of $D(T^{2})$ depends strongly on the $\Delta/k_{\mathrm{B}%
}T_{\mathrm{c}}$ ratio and it is also expected to be sensitive to the symmetry
of the superconducting energy gap.

Bearing in mind the anisotropic single gap behavior of elemental aluminum at
ambient pressure,\cite{Wells_PR_1970, Blackford_JLTP_1976, Kogure_JPSJ_1985}
the analysis of $B_{\mathrm{c}}(T,p)$ dependencies was performed by means of
the phenomenological model of Clem,\cite{Clem_AnnPhys_1966, Clem_PR_1967}
which considers effects of gap anisotropy on fundamental thermodynamic
quantities of a superconductor, and is correct in the weak anisotropy limit.
Strictly speaking, the analysis of temperature dependencies of various
thermodynamic quantities in an anisotropic superconductor requires the exact
knowledge of a momentum dependence of the superconducting energy gap, as well as
the shape of the Fermi surface. Using the spherical Fermi surface
approximation, Clem\cite{Clem_AnnPhys_1966} obtained the analytical
expressions for $B_{\mathrm{c}}(T)$ and $D(T)$ in the low- and the
high-temperature regimes in terms of the mean-squared anisotropy,
Eq.~\ref{eq:anisotropy-squared}.

\begin{figure}[tbh]
%\centering
\includegraphics[width=0.8\linewidth]{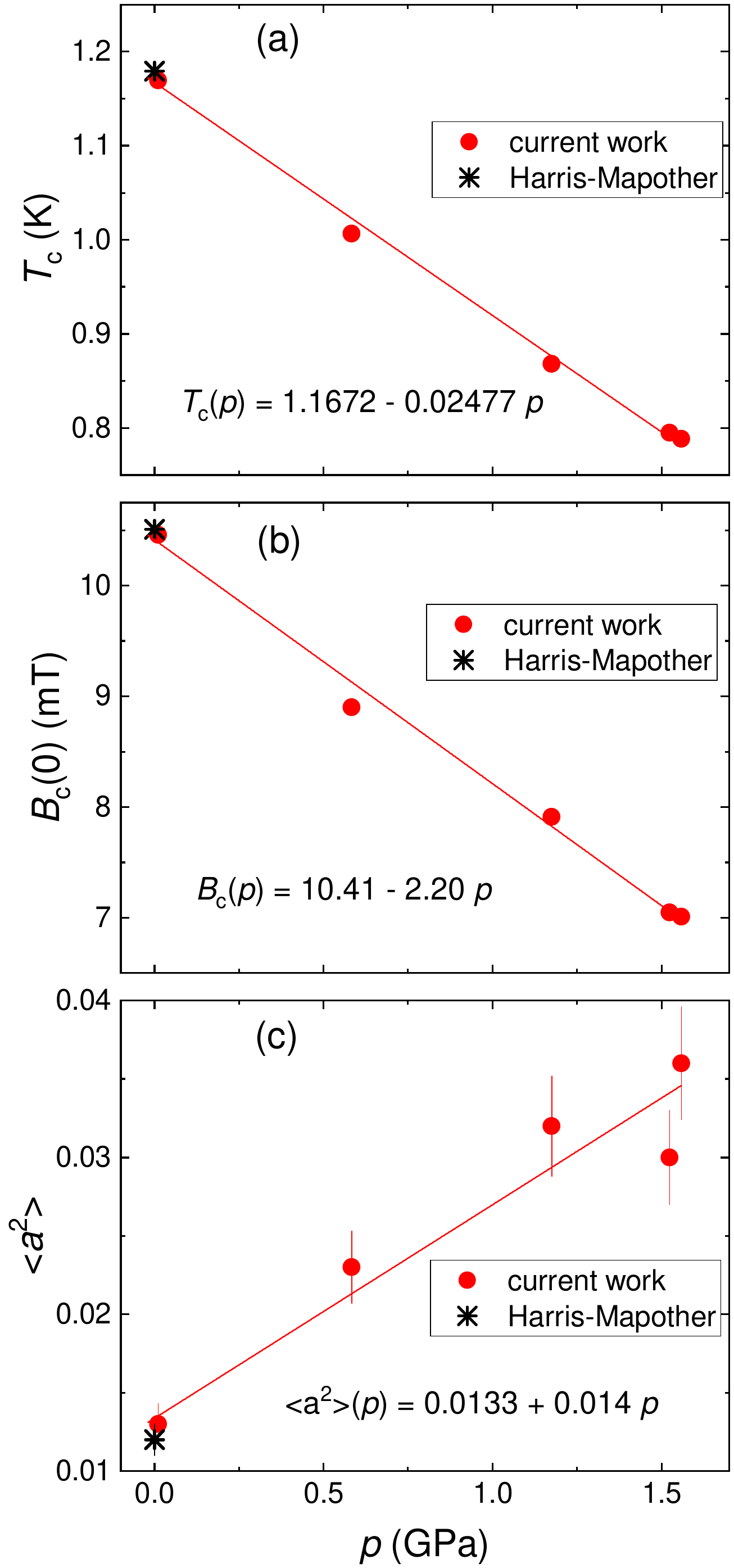}
%\vspace{-1.0cm}
\caption{Pressure dependence of: (a) the superconducting transition
temperature $T_{\mathrm{c}}$; (b) the zero-temperature value of the
thermodynamic critical field $B_\mathrm{c}(0)$; and (c) the mean-squared
anisotropy $\langle a^{2}\rangle$ of elemental aluminum. The quantities were
obtained from the analysis of $B_{\mathrm{c}}(T^{2},p)$ and $D(T^{2},p)$ data
within the framework of anisotropic model of Clem.\cite{Clem_AnnPhys_1966,
Clem_PR_1967} The circles are the data from the present study, and the
asterisks are obtained by analyzing the data of Harris and
Mapother.\cite{Harris_PR_1968} The lines are linear fits. }%
\label{fig:Parameters_vs_p}%
\end{figure}

Following Clem,\cite{Clem_AnnPhys_1966} in the low-temperature regime
($T/T_{\mathrm{c}}\lesssim0.3$):
\begin{equation}
b_{\mathrm{c}}(t)\simeq1-1.057(1+2\langle a^{2}\rangle)t^{2} -
0.559(1+4\langle a^{2}\rangle)t^{2} \label{eq:Bc_low-T}%
\end{equation}
and
\begin{equation}
D(t)\simeq-(0.057+2.11\langle a^{2}\rangle)t^{2}-0.559(1+4\langle a^{2}%
\rangle)t^{4}. \label{eq:D_low-T}%
\end{equation}
Here, the notations of the reduced temperature $t=T/T_{\mathrm{c}}$ and the
reduced field $b_{\mathrm{c}}(t)=B_{\mathrm{c}}(T)/B_{\mathrm{c}}(T=0)$ are used.

At higher temperatures ($0.7\lesssim T/T_{\mathrm{c}}\lesssim1$), the
equations for $b_{\mathrm{c}}(t)$ and $D(t)$ become:\cite{Clem_AnnPhys_1966}
\begin{align}
b_{\mathrm{c}}(t) &  \simeq1.7367(1-\langle a^{2}\rangle)(1-t)^{2}%
[1-(0.273-0.908\langle a^{2}\rangle)\nonumber\\
&  -(0.0949-0.037\langle a^{2}\rangle)(1+t)]
 \label{eq:Bc_high-T}
\end{align}
and
\begin{align}
D(t) &  \simeq-0.1317(1+6.6\langle a^{2}\rangle)(1-t^{2})\nonumber\\
&  +0.0986(1+3\langle a^{2}\rangle)(1-t^{2})^{2}\nonumber \\
&  +0.0287(1+6.15\langle a^{2}\rangle)(1-t^{2})^{3}.
 \label{eq:D_high-T}
\end{align}

\begin{figure}[tbh]
%\centering
\includegraphics[width=0.8\linewidth]{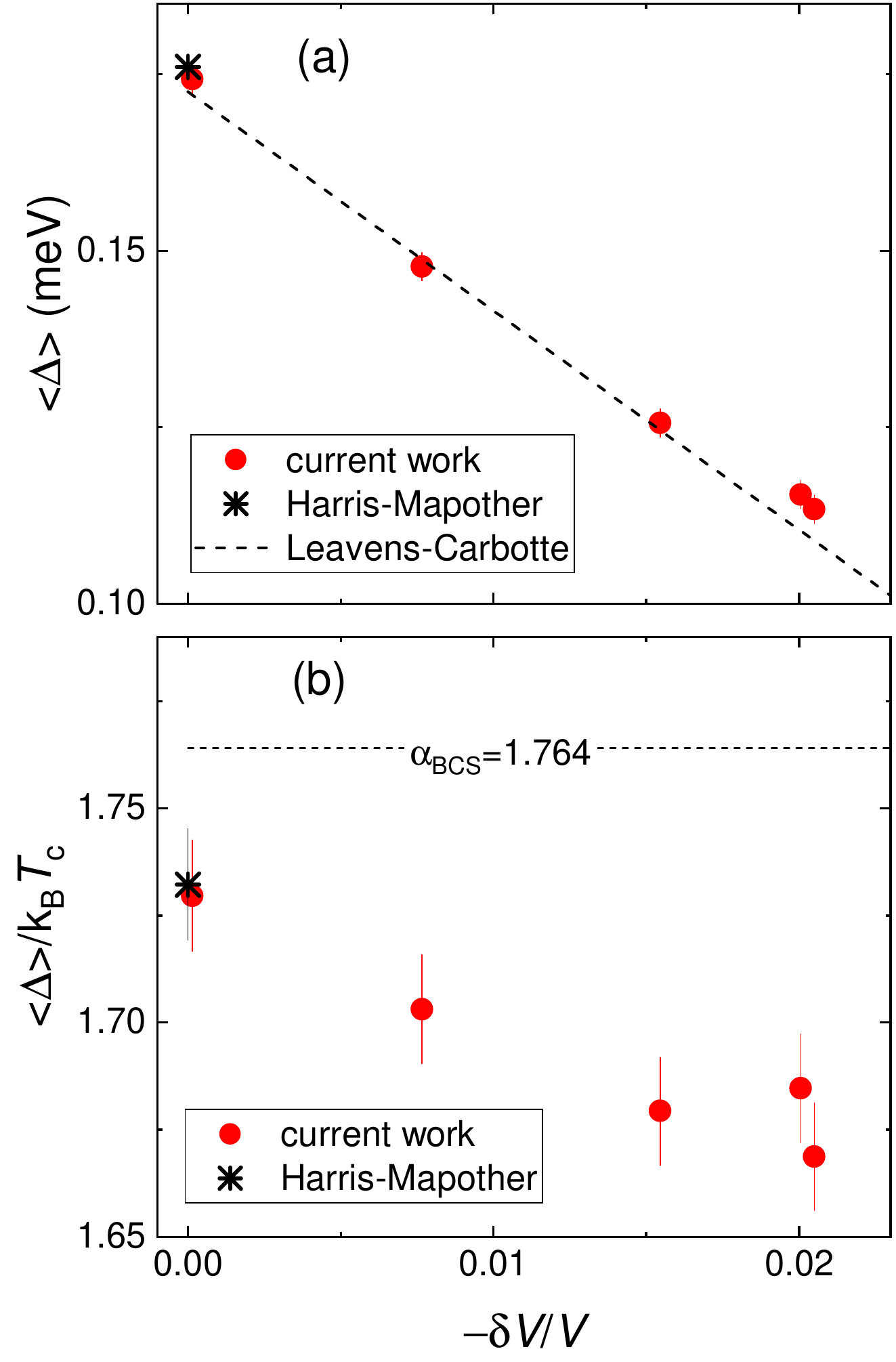}
%\vspace{-1.0cm}
\caption{(a) Dependence of the averaged gap value $\langle\Delta\rangle$ on
the relative volume change $-\delta V/V$. The dashed line is the theory
prediction of Leavens and Carbotte.\cite{Leavens_CanJPhys_1972} (b) Dependence
of $\langle\alpha\rangle=\langle\Delta\rangle/k_{\mathrm{B}} T_{\mathrm{c}}$
on $-\delta V/V$. The dashed line is the weak-coupling BCS number
$\alpha_{\mathrm{BCS}}\simeq1.764$. The closed circles correspond to the data
obtained within the present study. The black asterisks are parameters obtained
after analyzing the data of Harris and Mapother.\cite{Harris_PR_1968}}%
\label{fig:Parameters_vs_V}%
\end{figure}

The results of the fit of Eqs.~\ref{eq:Bc_low-T}, \ref{eq:Bc_high-T} and
Eqs.~\ref{eq:D_low-T}, \ref{eq:D_high-T} to the $B_{\mathrm{c}}(T^{2},p)$ and
$D(T^{2},p)$ data are presented in Figs.~\ref{fig:Bc_Deviation}~(a) and (b) by solid lines. The parameters obtained from the fits, namely the
superconducting transition temperature $T_{\mathrm{c}}$, the zero-temperature
value of the thermodynamic critical field $B_{\mathrm{c}}(0)$, and the
mean-squared anisotropy $\langle a^{2}\rangle$ are summarized in
Fig.~\ref{fig:Parameters_vs_p}. In addition to the results of the present
study, the experimental $B_{\mathrm{c}}(T,p=0)$ data of Harris and Mapother,
Ref.~\onlinecite{Harris_PR_1968}, were also reanalyzed (black asterisk symbols
at Figs.~\ref{fig:Parameters_vs_p} and \ref{fig:Parameters_vs_V}).

From the results presented in Fig.~\ref{fig:Parameters_vs_p} the following
four important points emerge:\newline(i) The fit parameters for an ambient
pressure TF-$\mu$SR data coincide with that obtained by reanalyzing the
results of magnetization data of Harris and Mapother.\cite{Harris_PR_1968}
\newline(ii) The linear fit of $T_{\mathrm{c}}(p)$ and $B_{\mathrm{c}}(0,p)$
dependencies result in $\mathrm{d}T_{\mathrm{c}}/\mathrm{d}p\simeq0.25$~K/GPa
and $\mathrm{d}B_{\mathrm{c}}(0,p)/\mathrm{d}p\simeq2.2$~mT/GPa, respectively,
which stays in a good agreement with the previously published
data.\cite{Boughton_book_1970, Palmy_Physica_1971, Levy_SSC_1964,
Gubser_PRL_1975}\newline(iii) The zero-pressure value of the mean-squared
anisotropy parameter was found to be $\langle a^{2}\rangle\simeq0.013$. This
value stays in agreement with the results of tunneling experiments of
Blackford,\cite{Blackford_JLTP_1976} and Kogure \textit{et al.}%
\cite{Kogure_JPSJ_1985} and the low-temperature specific heat measurements of
Cheeke \textit{et al.}\cite{Cheeke_JLTP_1973} reporting $\langle a^{2}%
\rangle\simeq0.009-0.01$. Markowitz and Kadanoff\cite{Markowitz_PhyRev_1959}
estimate $\langle a^{2}\rangle\simeq0.011$, by analyzing measurements of the
critical temperature as a function of impurity doping by Chanin \textit{et
al.}\cite{Chanin_PhysRev_1959} The theoretical predictions of Leavens and
Carbotte\cite{Leavens_AnnPhys_1972, Leavens_CanJPhys_1971} and Meza-Montes
\textit{et al.}\cite{Meza-Montes_JLTP_1988} gave $\langle a^{2}\rangle=0.0084$
and 0.015, respectively.\newline(iv) With the pressure increase from $p=0.0$
to $\simeq1.6$~GPa the mean-squared anisotropy parameter $\langle a^{2}%
\rangle$ almost triples, from $\simeq0.013$ to $\simeq0.035$ [see
Fig.~\ref{fig:Parameters_vs_p}~(c)]. Bearing in mind that within the model of
Clem\cite{Clem_AnnPhys_1966, Clem_PR_1967} the anisotropic gap behavior
follows:
\begin{equation}
\Delta(\mathbf{k})=\langle\Delta\rangle[1+a(\mathbf{k)],} \nonumber
\label{eq:anisotropic-gap}%
\end{equation}
this suggests that the ratio between the smallest and the biggest gap
[$\Delta(\mathbf{k})^{\mathrm{max}}/\Delta(\mathbf{k})^{\mathrm{min}}$]
increases from $\simeq1.25$ at ambient pressure to $\simeq2.2$ at $p\simeq
1.6$~GPa. Here, the simplest case (the two-gap case) with $a(\mathbf{k})=\pm
a_{0}$ and with the equal weight of $\langle\Delta\rangle(1\pm a_{0})$ gaps
was considered.

As the next step, we are going to compare the results of present study with
the calculations of Leavens and Carbotte.\cite{Leavens_CanJPhys_1972}
Figure~\ref{fig:Parameters_vs_V}~(a) compares the average gap values
$\langle\Delta\rangle$ as a function of the reduced volume $-\delta V/V$. From
the present data the value of $\langle\Delta\rangle$ was calculated
via:\cite{Clem_AnnPhys_1966}
\begin{equation}
\frac{\langle\Delta\rangle}{k_{\mathrm{B}}T_{\mathrm{c}}}=1.764\left(
1-\frac{3}{2}\langle a^{2}\rangle\right) \nonumber
 \label{eq:alpha_anisotropic}%
\end{equation}
and the relative volume change was obtained as $-\delta V/V=K\cdot p$
($K\simeq76$~GPa is the bulk modulus of aluminum,
Ref.~\onlinecite{Aluminium_Bulk_Modulus}). Figure~\ref{fig:Parameters_vs_V}%
~(a) points to a very good agreement between the theory and the experiment.
%The experimentally observed averaged gap values follow the theory prediction.

Figure~\ref{fig:Parameters_vs_V}~(b) shows the dependence of $\langle
\alpha\rangle=\langle\Delta\rangle/k_{\mathrm{B}}T_{\mathrm{c}}$ on the
relative volume change $-\delta V/V$. The results presented in the graph are
two-fold:
\newline(i) All the experimental points lie below the weak-coupling
BCS value $\alpha_{\mathrm{BCS}}\simeq1.764$. As already stays in the
introduction, it is not possible for isotropic superconductor to have the
coupling strength parameter smaller than the universal BCS number:
$\alpha\nless\alpha_{\mathrm{BCS}}$. This implies that the superconducting
energy gap in elemental aluminum \textit{must} be anisotropic and it
\textit{remains} anisotropic within the full pressure range studied here. Only in this particular case $\langle\alpha\rangle$ of aluminum may become smaller than $\alpha_{\mathrm{BCS}}$.
\newline(ii) The sample compression leads to continuous decrease of
$\langle\alpha\rangle$. Our experiments for aluminum indicate that the effect
is quite large. $\langle\alpha\rangle$ decreases by slightly more than 5\% for
a relative volume change of $\simeq2$\%.\newline
%Both this findings stay in agreement with the theory predictions of Clem, Refs.~\onlinecite{Clem_AnnPhys_1966, Clem_PR_1967} and Leavens and Carbotte. Ref.~\onlinecite{Leavens_CanJPhys_1972}.

Leavens and Carbotte, Ref.~\onlinecite{Leavens_CanJPhys_1972}, provided a
simple explanation for the pressure dependence of the superconducting gap
anisotropy. Following
Refs.~\onlinecite{Clem_AnnPhys_1966, Leavens_CanJPhys_1972}, the dominant
source of the gap anisotropy is the anisotropy of the phonon mediated
electron-electron interaction. When the metal is subjected to a hydrostatic
pressure, the Fermi sphere and the Brillouin zone scale together, so that the
shape of the complicated surface generated for the initial point $(\theta
,\phi)$ does not change.
%while it size changes, of course.
It is expected, therefore, that even though the absolute value of the
electron-phonon coupling constant $\lambda_{\mathrm{e-p}}$ changes as a
function of pressure, the anisotropy of the phonon induced electron-electron
interaction remains unaltered. Bearing in mind that the Coulomb
pseudopotential parameter $\mu^{\ast}$ changes very slowly with pressure as
compared to the phonon-mediated interaction, the anisotropy in the total
interaction increases. This said, full density functional theory calculations
of the anisotropic Eliashberg function and \textbf{k}-dependent order
parameter in Al are highly desirable.

The above arguments would also imply, that smaller is the ratio of the
electron-phonon interaction to the strength of the Coulomb interaction, the
faster will be the increase of the energy gap anisotropy with pressure. In
strong coupling superconductors, $\lambda_{\mathrm{e-p}}$ is about 10 times
higher than $\mu^{\ast}\sim0.1-0.15$, which is only weakly dependent on the
material. Note that the size of $\mu^{\ast}$ is fixed by the Tolmachev
logarithm:\cite{Tolmachev_DoklAkadNauk_1961} $\mu^{\ast}\approx\mu/[1+\mu
\ln(\theta_{D}/T_{\mathrm{c}})\approx1/\ln(\theta_{D}/T_{\mathrm{c}})$. It
might be expected, therefore, that the variation of the electron-phonon
interaction as a function of pressure in strongly coupled superconductors will
lead to a small pressure effect on the gap anisotropy. In a weakly coupled
superconductors, however, $\lambda_{\mathrm{e-p}}$ is smaller and the pressure
dependence of the superconducting gap anisotropy is expected to be more
pronounced. This is definitively the case of aluminum, where $\lambda
_{\mathrm{e-p}}\sim4\mu^{\ast}$.

To conclude, the pressure dependence of the thermodynamic critical field
$B_{\mathrm{c}}$ in elemental aluminum was studied by means of the muon-spin
rotation/relaxation. Pressure was found to enhance the deviation of
$B_{\mathrm{c}}(T)$ from the parabolic behavior, thus suggesting the weakening
of the reduced gap $\langle\alpha\rangle=\langle\Delta\rangle/k_{\mathrm{B}%
}T_{\mathrm{c}}$. The analysis of the experimental data within the anisotropic
gap model of Clem,\cite{Clem_AnnPhys_1966, Clem_PR_1967} suggests that with
the pressure increase from 0.0 to $\simeq1.6$~GPa the mean-squared anisotropy
value $\langle a^{2}\rangle$ changes from $\simeq0.013$ to $\simeq0.035$,
which, in turn, leads to a change of $\langle\alpha\rangle$ from 1.73 to 1.67.

This work was performed at the Swiss Muon Source (S$\mu$S), Paul Scherrer
Institute (PSI, Switzerland). RK acknowledges the technical support of
Matthias Elender, Ritu Gupta and Debarchan Das. The authors acknowledge
helpful discussions with Rafael Fernandes.

\end{document}